\documentclass[preprint,showpacs,preprintnumbers,amsmath,amssymb,superscriptaddress,12pt,graphicx]{revtex4}

\usepackage{graphicx}
\usepackage{dcolumn}
\usepackage{bm}
\usepackage{amssymb}

\renewcommand{\L}{{\cal L}}

\newcommand{\be}{\begin{equation}}
\newcommand{\ee}{\end{equation}}
\newcommand{\bee}{\begin{eqnarray}}
\newcommand{\een}{\end{eqnarray}}
\newcommand{\ba}{\begin{eqnarray}}
\newcommand{\ea}{\end{eqnarray}}

\begin{document}

\title{Statefinder diagnostic for Yang-Mills dark energy model  }

\author{Wen Zhao}
\email{wzhao7@mail.ustc.edu.cn}\affiliation{Department of Applied Physics, Zhejiang University of Technology,  Hangzhou, Zhejiang}\affiliation {Department of Physics and Astronomy, Cardiff University, Cardiff, United Kingdom} \affiliation{Wales Institute of Mathematical and Computational Science, Swansea, United Kingdom}

\date{\today}


\begin{abstract}
We study the statefinder parameters in the Yang-Mills
condensate dark energy models, and find that the evolving
trajectories of these models are different from those of other
dark energy models. We also define two eigenfunctions of the
Yang-Mills condensate dark energy models. The values of these
eigenfunctions are quite close to zero if the equation-of-state of
the Yang-Mills condensate is not far from $-1$, which can be used
to simply differentiate between the Yang-Mills condensate models
and other dark energy models.
\end{abstract}



\maketitle


\section{Introduction}

Physicists and astronomers begin to consider the dark energy
cosmology seriously and to explore the nature of dark energy
actively since the expansion of our universe is proven to be
accelerating at present time by the Type Ia supernovae
observations\cite{sn}. The analysis of cosmological observations,
in particular of the WMAP (Wilkinson Microwave Anisotropy Probe)
experiment\cite{map}, indicates that dark energy occupies about
$70\%$ of the total energy of our universe, and dark matter about
$26\%$. The accelerated expansion of the present universe is
attributed to that dark energy is an exotic component with
negative pressure, and the simplest candidate for dark energy is
the cosmological constant. However, two difficulties arise from
this scenarios, namely the fine-tuning problem and the cosmic
coincidence problem\cite{problem}. So the dynamical models are
considered by a number of authors, such as the quintessence,
phantom, k-essence, and quintom models\cite{models}. The effective
Yang-Mills  condensate (YMC) as a kind of candidate for dark
energy has been detailed discussed in the
Refs.\cite{z,Zhang,zhao}. The effective lagrangian up to 1-loop
order is \cite{pagels, adler}
 \be
 \L_{eff}=\frac{b}{2}F\ln\left|\frac{F}{e\kappa^2}\right|, \label{L}
 \ee
where  $b=11N/24\pi^2$ for the generic gauge group $SU(N)$ is the
Callan-Symanzik coefficient\cite{Pol}.
$F=-(1/2)F^a_{\mu\nu}F^{a\mu\nu}$ plays the role of the order
parameter of the YMC, $\kappa$ is the renormalization scale, the
only model parameter. The attractive features of this effective
lagrangian include the gauge invariance, the Lorentz invariance,
the correct trace anomaly, and the asymptotic
freedom\cite{pagels}. The effective YMC was firstly put into the
expanding Friedmann-Robertson-Walker (FRW) spacetime to study
inflationary expansion \cite{z} and the dark energy \cite{Zhang}.
We work in a spatially flat FRW spacetime with a metric
 \be
 ds^2=a^2(\tau)(d\tau^2-\delta_{ij}dx^idx^j),\label{me}
 \ee
where $\tau=\int(a_0/a)dt$ is the conformal time. For simplicity
we study the $SU(2)$ group. Compared with the scalar field, the YM
field is the indispensable cornerstone to particle physics and the
gauge bosons have been observed. There is no room for adjusting
the form of effective YM lagrangian as it is predicted by quantum
corrections according to field theory. In the previous works, we
have deeply investigated the 1-order YMC models and found
attractive features: \emph{a.} this dark energy can naturally get
the equation-of-state (EOS) of $w>-1$ and $w<-1$\cite{zhao}, which
is different from the scalar quintessence models; \emph{b.} in the
free field models, with the expansion of the universe, the EOS of
the YMC naturally turns to the critical state of
$w=-1$\cite{zhao}, consistent to the observations; \emph{c.} the
cosmic coincidence problem is naturally avoided in the
models\cite{zhao}; \emph{d.} the EOS of the dark energy can cross
$-1$ in the double-field models or coupled models\cite{zhao};
\emph{e.} the big rip is naturally avoided in the
models\cite{zhao}; \emph{f.} the magnetic component of the YMC
naturally decreases to zero with the expansion of the
universe\cite{zhao}.

In the recent paper\cite{xia}, the authors have detailed discussed
the 2-loop YMC dark energy. In this model, the effective
lagrangian is
 \be
 {\mathcal{L}}_{eff}=\frac{b}{2}F\left[\ln\left|\frac{F}{e\kappa^2}\right|
 +\eta\ln\left|\ln\left|\frac{F}{e\kappa^2}\right|+\delta\right|~\right]
 \ee
where $\eta\simeq0.84$ and the dimensionless constant $\delta$ is
a parameter representing higher order corrections. In this 2-loop
model, the cosmic coincidence problem is also naturally avoided.
This feature is same with the 1-loop models. From the Einstein
equation\cite{xia}, we can easily find that, in the free field
models, the late time attractor exists, which satisfies the
relation
 \be
 \beta_c=-\eta\left(\log|\beta_c-1+\delta|+\frac{1}{\beta_c-1+\delta}\right),
 \ee
where $\beta=\ln|F/\kappa^2|$. It is easily found that $w=-1$,
when $\beta=\beta_c$. So in these models, the EOS of YMC also
naturally turns to the critical state $w=-1$. However, the
cosmological constant crossing is also naturally realized, if
considering the interaction between the YMC and matter \cite{xia}.
The discussion in \cite{xia} shows that, although the 2-loop
models are much more general and complicated than the 1-loop
model, the most physics features  are not unchanged.

In this paper, we only consider the 1-loop models. As in previous
works, we only consider the electric case with $B^2 \equiv0 $. The
energy density and pressure of YMC are given by
 \be
 \rho_y=\frac{E^2}{2}\left(\epsilon+b\right),
 ~~~~p_y=\frac{E^2}{2}\left(\frac{\epsilon}{3}-b\right),
 \ee
where the dielectric constant is
 \be
 \epsilon=b\ln\left|\frac{F}{\kappa^2}\right|.\label{epsilon}
 \ee
and the EOS is obtained
 \be
 w=\frac{p_y}{\rho_y}= \frac{\beta-3}{3\beta+3},\label{13}
 \ee
where $\beta\equiv\epsilon/b$. At the critical point with the
condensate order parameter $F=\kappa^2$, one has $\beta=0$ and
$w=-1$. Around this critical point, $F< \kappa^2$ gives $\beta<0$
and $w<-1$, and $F> \kappa^2$ gives $\beta>0$ and $w>-1$. So in
the YMC model, EOS of $w >-1$ and $w<-1$ can be naturally
realized. When $\beta\gg1$, the YM field has a state of $w=1/3$,
becoming a radiation component.

In the free field models, the effective YM equations are
 \be
 \partial_{\mu}(a^4\epsilon~
 F^{a\mu\nu})+f^{abc}A_{\mu}^{b}(a^4\epsilon~F^{c\mu\nu})=0,
 \label{F1}
 \ee
which reduces to\cite{zhao}
 \be
 \partial_{\tau}(a^2\epsilon E)=0.
 \ee
At the critical point ($\epsilon=0$), this equation is an
identity. When $\epsilon\neq 0$, this equation has an exact
solution \cite{zhao}:
 \be
 \beta~ e^{\beta/2}\propto a^{-2},\label{16}
 \ee
where the coefficient of proportionality depends on the initial
condition. For a fixed initial condition, we can obtain the
evolution of the EOS of the YMC by using the YM equation
(\ref{16}). In the previous works, we found the free YMC can be
separated into two kinds, the quintessence-like or phantom-like,
which only depends on the choice of the initial condition. In
order to differentiate between the YMC dark energy models and
other models, a sensitive and robust diagnostic for dark energy
models is needed. For this purpose a diagnostic proposal that
makes use of parameter pair $\{r,s\}$, the so-called
``statefinder", was introduced by Sahni et al.\cite{sahni}. The
statefinder probes the expansion dynamics of the universe through
higher derivatives of the expansion factor $\stackrel{...}{a}$ and
is a natural companion to the deceleration parameter $q$ which
depends upon $\ddot a$. The statefinder pair $\{r,s\}$ is defined:
\begin{equation}
r\equiv
\frac{\stackrel{...}{a}}{aH^3},~~~~s\equiv\frac{r-1}{3(q-1/2)}~.\label{rs}
\end{equation} The statefinder
is a ``geometrical'' diagnostic in the sense that it depends upon
the expansion factor and hence upon the metric describing
space-time.

Trajectories in the $s-r$ plane corresponding to different
cosmological models exhibit qualitatively different behaviors. The
spatially flat LCDM (cosmological constant $\Lambda$ with cold
dark matter) scenario corresponds to a fixed point in the diagram
$\{s,r\} = \{ 0,1\}$. Departure of a given dark energy model from
this fixed point provides a good way of establishing the
``distance'' of this model from LCDM \cite{sahni,alam}. As
demonstrated in Refs. \cite{sahni,alam,gorini} the statefinder can
successfully differentiate between a wide variety of dark energy
models including the cosmological constant, quintessence, the
Chaplygin gas, braneworld models and interacting dark energy
models. We can clearly identify the ``distance'' from a given dark
energy model to the LCDM scenatio by using the $r(s)$ evolution
diagram.

The current location of the parameters $s$ and $r$ in these
diagrams can be calculated in models, and on the other hand it can
also be extracted from data coming from SNAP (SuperNovae
Acceleration Probe) type experiments \cite{snap}. Therefore, the
statefinder diagnostic combined with future SNAP observations may
possibly be used to discriminate between different dark energy
models. In this letter we apply the statefinder diagnostic to YMC
dark energy models.


\section{Statefinder for YMC dark energy}

The statefinder parameters $r$ and $s$ in (\ref{rs}) can be
rewritten as
\begin{equation}
r=1+{9\over 2}w(1+w)\Omega_y-{3\over 2}w'\Omega_y~,
\end{equation}
\begin{equation}
s=1+w-{1\over 3}{w'\over w}~.
\end{equation}
where $w$ is the EOS of the YMC, and $\Omega_y$ is the fractional
energy density of YMC. A prime denotes derivation with respect to
the e-folding time $N\equiv\ln a$. From the previous discussion,
we know
 \be
 w=\frac{\beta-3}{3\beta+3},~~{\rm
 and}~~w'=\beta'\frac{dw}{d\beta}. \label{wwp}
 \ee
So the statefinder for YMC only depends on the evolution of
parameter $\beta$, which can be exactly determined by the YM
equation (\ref{16}). From the YM equation, we obtain that
 \be
 \beta'=\frac{-4\beta}{2+\beta},~~{\rm and}~~
 w'=\frac{-16\beta}{3(1+\beta)^2(2+\beta)},
 \ee
and the statefinder parameters become
 \be
 r=\frac{2+(3-4\Omega_y)\beta+(1+2\Omega_y)\beta^2}{2+3\beta+\beta^2}\label{r},
 \ee
 \be
 s=\frac{4\beta(\beta-2)}{3(\beta^2-\beta-6)}\label{s}.
 \ee
The deceleration parameter is also obtained
 \be
 q=\frac{1+\beta-3\Omega_y+\beta\Omega_y}{2+2\beta}\label{q}.
 \ee

We first consider the case with $w>-1$, quintessence-like, where
$\beta>0$ is kept for all time. In the very early universe with
$\beta\gg1$ and $\Omega_y\rightarrow0$\cite{zhao}, we obtain
 \be
 r\rightarrow1, s\rightarrow\frac{4}{3},~{\rm
 and}~q\rightarrow\frac{1}{2},
 \ee
which is independently of the initial condition, and obviously
different from the SCDM (standard cold dark matter) model with
$(r,s,q)=\left(1,1,1/2\right)$. In the later stage of the universe
with $\beta\rightarrow0$ and $\Omega_y\rightarrow1$\cite{zhao},we
have
 \be
 r\rightarrow1, s\rightarrow0,~{\rm
 and}~q\rightarrow-1.
 \ee
The universe approaches an exact de Sitter expansion, which is
same with the late stage of the LCDM model. We also notice that
the value of $s$ is infinite when $\beta=3$, where the EOS of the
YMC is $w=0$. This is also a character of the YMC models.

If the YMC is phantom-like with $w<-1$ and $\beta<0$. In the late
stage of the universe, we have $\beta\rightarrow0$ and
$\Omega_y\rightarrow1$\cite{zhao}, which follows that
 \be
 r\rightarrow1, s\rightarrow0,~{\rm
 and}~q\rightarrow-1.
 \ee
which is same with the quintessence-like case. Here we should
point out that, in the very early universe with $a\rightarrow0$,
the YM kinetic equation (\ref{16}) has no solution for the case
with $w_0<-1$, where $w_0$ is the present EOS of the YMC. So the
free YMC is not applied in the very early universe, where the
interaction between the YMC and matter\cite{zhao,xia}, or the
phase transition of the YMC must be considered.

The evolution of the parameter $\beta$ and $\Omega_y$ can be
obtained by Eq.(\ref{F1}) for a fixed initial condition. In Fig.1,
we show the evolution of the statefinder pair ${s,r}$, where the
initial conditions of YMC are $w_0=-1.1,-0.9,-0.8$, respectively,
and the present fraction energy density of YMC is $\Omega_y=0.7$.
It can be found that the trajectories of these models never cross
the LCDM fixed point. However, with the expansion of the universe,
they will approach this fixed point, which is independent of the
choice of the initial condition. The only difference for the
quintessence-like and phantom-like cases is the direction of the
trajectories, when the models approach the fixed point. The
coordinate of today's points are
$(-0.033,1.113),(0.034,0.903),(0.070,0.824)$, respectively, thus
the distance from these models to the LCDM can be easily
identified in this diagram.

We also plot the evolution trajectories of statefinder pair
${r,q}$ in Fig.2.


\section{ Statefinder of first order}

Cosmological observations show that the EOS of the dark energy is
closer to $-1$. In the YMC dark energy models, from the expression
of EOS of YMC in (\ref{13}), we find that $w\rightarrow-1$ follows
that $|\beta|\ll1$. So we can Taylor expand the EOS and the
statefinder of the YMC with parameter $\beta$ at the critical
state of $w=-1$. Keeping the first order of the smaller quantity
$\beta$, we can rewrite the EOS of the YMC as
 \be
 w=-1+\frac{4}{3}\beta+O(\beta^2),~~{\rm and}~~w'=-\frac{8}{3}\beta+O(\beta^2).
 \ee
From the expressions in (\ref{r}),(\ref{s}) and (\ref{q}), we
obtain
 \be
 r=1-2\Omega_y\beta+O(\beta^2),~~
 s=\frac{4}{9}\beta+O(\beta^2)\label{rs1}
 \ee
and the deceleration parameter
 \be
 q=\left(\frac{1}{2}-\frac{3\Omega_y}{2}\right)+2\Omega_y\beta+O(\beta^2).
 \ee
These functions only depend on the quantities $\beta$ and
$\Omega_y$, which are all determined by the initial condition, and
the initial condition of the YMC directly relates to the present
EOS of the YMC.

In order to differentiate between the YMC dark energy models and
other models, such as the quintessence, phantom, k-essence, or
chaplygin gas, we can define an eigenfunction of first order for
the YMC models
 \be
 \xi_1=\frac{2}{9}\frac{r-1}{\Omega_y}+s.
 \ee
From the expressions of (\ref{rs1}), we find that the value of
this eigenfunction is $\xi_1=0+O(\beta^2)$, which is independent
of the initial condition of the YM dark energy models. It is easy
to find that this feature is not right for other dark energy
models. So we can differentiate the YMC dark energy models from
other models by the observable quantity $\xi_1$.


\section{Statefinder of second order}

We also can expand the EOS and statefinder of YMC to the second
order of $\beta$. From the expressions in (\ref{wwp}), we obtain
 \be
 w=-1+\frac{4}{3}\beta-\frac{4}{3}\beta^2+O(\beta^3),~~{\rm and}~~w'=-\frac{8}{3}\beta+\frac{20}{3}\beta^2+O(\beta^3).
 \ee
The statefinder parameters are
 \be
 r=1-2\Omega_y\beta+4\Omega_y\beta^2+O(\beta^3),
 \ee
 \be
 s=\frac{4}{9}\beta-\frac{8}{27}\beta^2+O(\beta^3),
 \ee
and the deceleration parameter is
 \be
 q=\left(\frac{1}{2}-\frac{3\Omega_y}{2}\right)+2\Omega_y\beta-2\Omega_y\beta^2+O(\beta^3).
 \ee

From these expressions, we can also define an eigenfunction of
second order
 \be
 \xi_2=-\frac{8}{27}\frac{r-1}{\Omega_y}+\frac{512}{9}(w+1)-44s.
 \ee
It is easy to find that the value of this eigenfunction
$\xi_2=0+O(\beta^3)$. In Fig.3, we plot the evolution of the
eigenfunctions $\xi_1$ and $\xi_2$ in the different YMC dark
energy models. We find that, in all these models, the values of
$\xi_1$ and $\xi_2$ are all very closer to $0$ if the EOS of the
YMC is not very far from $-1$. From this figure, we also find that
the value of $\xi_2$ is much more closer to $0$ than which of
$\xi_1$. The former is a more effective function for affirming the
YMC dark energy models. Of course, in the LCDM models, the values
of $\xi_1$ and $\xi_2$ are all exact $0$, so it is difficult to
differentiate between the YMC dark energy model and LCDM model.


\section{Summary}

In summary, we have investigated the statefinder of the YMC dark
energy models in this letter. We analyze two cases of the models,
the quintessence-like case and the phantom-like case, and perform
a statefinder diagnostic to both cases.  It is shown that the
evolving trajectory of this scenario in the $s-r$ plane is quite
different from those of other models. We also define two
eigenfunctions of YMC dark energy model. If the EOS of the YMC is
not far from $-1$, the values of the eigenfunctions are very
closer to $0$, which can be used to simply differentiate between
the YMC and other dark energy models.


\section*{Acknowledgements}
The author thanks the referee for help
discussions. This work is supported by CNSF No. 10703005 and
10775119, the Research Funds Launched in ZJUT No.109001729.


\newpage

 \begin{figure}
 \centerline{\includegraphics[width=15cm]{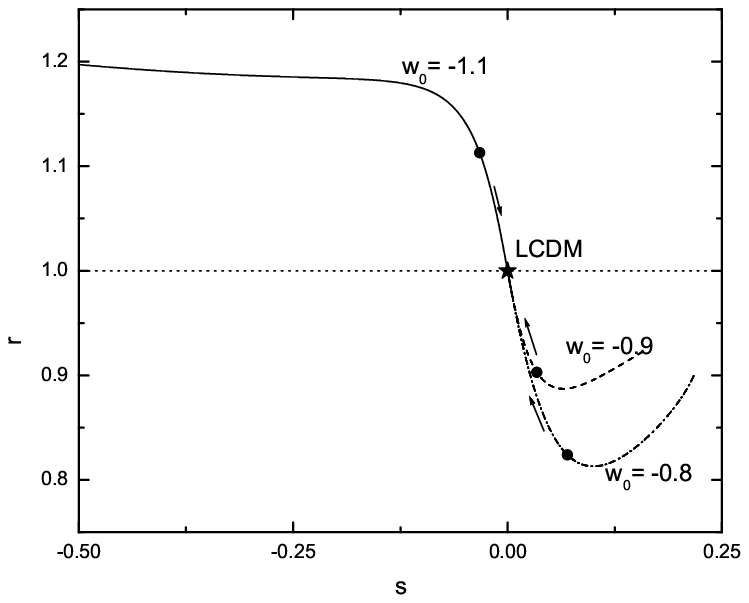}}
 \caption{\small The $s-r$ diagram of the YMC dark energy models. Dots locate the current
 values of the statefinder pair $\{s,r\}$, and the arrows denote the evolution direction of the
 statefinders with expansion of the universe.   }
 \end{figure}

 \begin{figure}
 \centerline{\includegraphics[width=15cm]{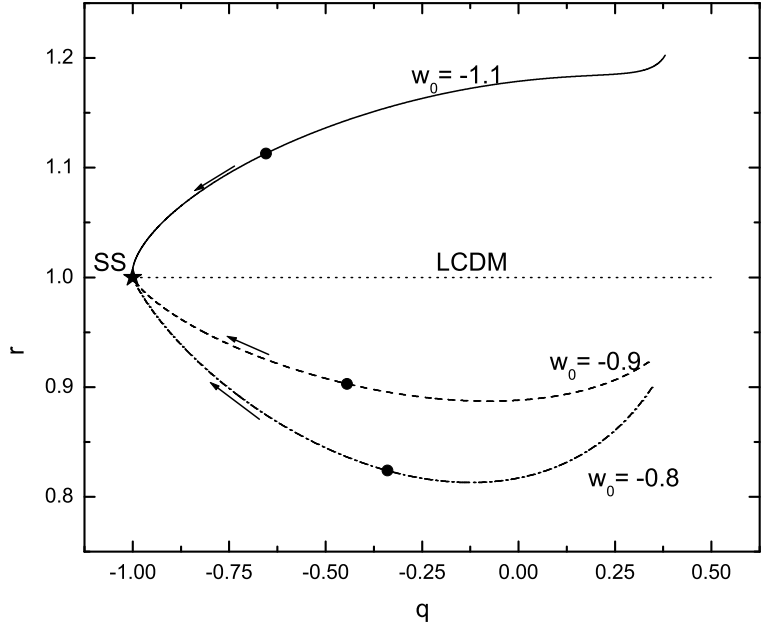}}
 \caption{\small The $q-r$ diagram of the YMC dark energy models. Dots locate the current
 values of the statefinder pair $\{q,r\}$, and the arrows denote the evolution direction of the
 statefinders with expansion of the universe. The point of $(-1,1)$ corresponds to the steady state models
 (SS) - the de Sitter expansion.}
 \end{figure}

 \begin{figure}
 \centerline{\includegraphics[width=15cm]{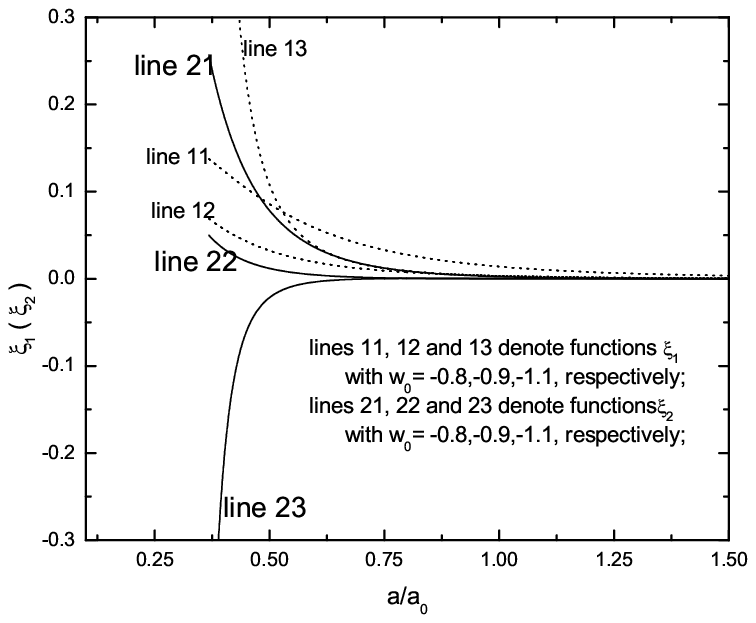}}
 \caption{\small The evolution of the eigenfunctions $\xi_1$ and $\xi_2$ with the scale factor $a$.}
 \end{figure}


\baselineskip=12truept

\begin{thebibliography}{99}


\bibitem{sn}
A.G.Riess et al., Astron.J. {\bf 116}, 1009 (1998);

S.Perlmutter et al., Astrophys.J. {\bf 517}, 565 (1999);

J.L.Tonry et al., Astrophys.J. {\bf 594}, 1 (2003);

R.A.Knop et al., Astrophys.J. {\bf 598}, 102 (2003);


\bibitem{map}
C.L.Bennett et al., Astrophys.J.Suppl. {\bf 148}, 1 (2003);

D.N.Spergel et al., Astrophys.J.Suppl. {\bf 148}, 175 (2003);

D.N.Spergel et al., Astrophys.J.Suppl. {\bf 170}, 377 (2007);



\bibitem{problem}
P.Steinhardt, in `Critical Problems in Physics' ed. by V.L.Fitch
and D.R.Marlow (Princeton U.Press, 1997);

R.H.Dicke and P.J.E.Peebles, in `General Relativity: An Einstein
Centenary Survey', ed. by S.W.Hawking \& W.Israel (Cambridge U.
Press, 1979);

E.J.Copeland, M.Sami and S.Tsujikawa, Int.J.Mod.Phys.D {\bf 15},
1753 (2006);



\bibitem{models}
C.Wetterich, Nucl.Phys.B {\bf 302}, 668 (1988);~Astron.Astrophys.
{\bf 301}, 321 (1995);

B.Ratra and P.J.E.Peebles, Phys.Rev.D {\bf 37}, 3406 (1988);

R.R.Caldwell, Phys.Lett.B {\bf 545}, 23 (2002);

S.M.Carroll, M.Hoffman and M.Trodden, Phys.Rev.D {\bf 68}, 023509
(2003);

W.Zhao, Chinese Physics {\bf 16}, 2830 (2007);

C.Armendariz-Picon, T.Damour and V.Mukhanov, Phys.Lett.B {\bf
458}, 209 (1999);

B.Feng, X.L.Wang and X.M.Zhang, Phys.Lett.B {\bf 607}, 35 (2005);

Z.K.Guo, Y.S.Piao, X.M.Zhang and Y.Z~Zhang, Phys.Lett.B {\bf 608},
177 (2005);

H.Wei, R.G.Cai and D.F.Zeng, Class.Quant.Grav. {\bf 22}, 3189
(2005);

W.Zhao and Y.Zhang, Phys.Rev.D {\bf 73}, 123509 (2006);

Y.F.Cai, H.Li, Y.S.Piao and X.M.Zhang, Phys.Lett.B {\bf 646} 141
(2007);

W.Zhao, Phys.Lett.B {\bf 655} 97 (2007);

Y.F.Cai, T.T.Qiu, R.Brandenberger, Y.S.Piao, and X.M.Zhang,
arXiv:0711.2187;

\bibitem{z} Y.Zhang, Phys.Lett.B {\bf 340}, 18 (1994); Chin.Phys.Lett.{\bf 14}, 237 (1997);


\bibitem{Zhang}
Y.Zhang, Gen.Rel.Grav. {\bf 34}, 2155 (2002);~{\bf 35}, 689
(2003);~Chin.Phys.Lett.{\bf 20}, 1899 (2003);

\bibitem{zhao}
W.Zhao and Y.Zhang, Class.Quant.Grav. {\bf 23}, 3405 (2006);

W.Zhao and Y.Zhang, Phys.Lett.B {\bf 690}, 64 (2006);

W.Zhao and D.H.Xu, Int.J.Mod.Phys.D accepted
(arXiv:gr-qc/0701136);

Y.Zhang, T.Y.Xia and W.Zhao, Class.Quant.Grav. {\bf 24}, 3309
(2007);


\bibitem{pagels}
H.Pagels and E.Tomboulis, Nucl.Phys.B {\bf 143}, 485 (1978);

\bibitem{adler}
S.Adler, Phys.Rev.D {\bf 23}, 2905 (1981);~Nucl.Phys.B {\bf 217},
3881 (1983);



\bibitem{Pol}
H.Politzer, Phys.Rev.Lett. {\bf 30}, 1346 (1973);

D.J.Gross and F.Wilzcek, Phys.Rev.Lett. {\bf 30}, 1343 (1973);


%
%

\bibitem{xia}
T.Y.Xia and Y.Zhang, Phys.Lett.B {\bf 656}, 19 (2007);


\bibitem{sahni} V.Sahni, T.D.Saini, A.A.Starobinsky and
U.Alam, JETP Lett. {\bf 77}, 201 (2003);

\bibitem{alam} U.Alam, V.Sahni, T.D.Saini and
A.A.Starobinsky, Mon.Not.Roy.ast.Soc.{\bf 344}, 1057 (2003);

\bibitem{gorini} V.Gorini, A.Kamenshchik and U.Moschella, Phys.Rev.D {\bf 67}, 063509 (2003);

X.Zhang, Phys.Lett.B {\bf 611}, 1 (2005);

X.Zhang, Int.J.Mod.Phys.D {\bf 14}, 1597 (2005);

W.X.Wu and H.W.Yu, Int.J.Mod.Phys.D {\bf 14}, 1873 (2005);

B.R.Chang, H.Y.Liu, L.X.Xu, C.W.Zhang and Y.L.Ping, JCAP {\bf
0701}, 016 (2007);

\bibitem{snap}
SNAP Collaboration, astro-ph/0507458, astro-ph/0507459.




\end{thebibliography}
\end{document}